\documentclass{amsart}
\usepackage{amscd}
\usepackage{amssymb}
\DeclareFontFamily{OMS}{cmsy}{%
\fontdimen16\font=3pt
\fontdimen17\font=3pt}
\def\dj{d\kern-.30em\raise1.25ex\vbox{\hrule width .3em height .03em}}
\def\Dj{D\rlap{\kern-.70em\raise0.75ex
\vbox{\hrule width .3em height .03em}}}

\makeatletter
\renewcommand{\subsection}{\@startsection{subsection}{2}{\z@}%
{\baselineskip}{0.5\baselineskip}{\bfseries}}
\def\l@section{\def\@tocpagenum##1{\hss{\bfseries ##1}}%
\@tocline{1}{8pt}{0pc}{}{\bfseries}}
\def\l@subsection{\def\@tocpagenum##1{}
\@tocline{2}{2pt}{2pc}{2pc}{}}
\makeatother
\def\bla#1{$(${\it #1\/{}}$)$}

\renewcommand{\thepage}{\ifnum\value{page}=1 \else\arabic{page}\fi}
\begin{document} 
 
\title[Introduction to Subquantum Mechanics]{Physics Beyond the Limits
of Uncertainty Relations}

\author{micho {\Dj}UR{\Dj}EVICH}

\address{
Instituto de Matematicas, UNAM, 
Area de la Investigacion Cientifica, 
Circuito Exterior, Ciudad Universitaria,
M\'exico DF, CP 04510, MEXICO
}
\email{ micho@matem.unam.mx\newline\indent\indent http://www.matem.unam.mx/{\~\/}micho}
\maketitle

\begin{flushright}{\it
To see the world in a grain of sand,\\
And a heaven in a wild flower;\\
Hold infinity in the palm of your hand,\\
And eternity in an hour.}\footnote{\hfill William Blake, 
Auguries of Innocence}
\end{flushright}

\section{Introduction}
Quantum mechanics is a physics of {\it microworld.} Its principal aim is to provide a 
mathematically coherent picture of physical reality at the deepest possible 
experimentally accessible level--{\it  the quantum level.} This means understanding 
phenomenas involving elementary particles and quantas of interactions, 
and the study of the internal structure of matter and fields.  

Quantum world is very different from the picture given by {\it classical mechanics},
which is a {\it macroworld} physics.

One of the principal purely quantum phenomenas is
{\it  complementarity}. If we consider all possible {\it  properties/attributes} of a given 
quantum system then it turns out that for every state of the system there exists 
an infinite collection of properties that are {\it ontically inapplicable} 
to the system in this particular state. This means that in such situations 
an attribute from the mentioned collection does not hold, and also 
its negation does not hold. 

This is {\it very different} from classical mechanics, where every possible attribute of a 
physical system has a definitive ($0$ or $1$) value in every state. In other words, either
the attribure holds or its negation holds in a given state of the system. 

All the properties of a classical system form a Boolean algebra. All possible attributes of
a quantum system form an essentially different structure---a non-distributive lattice. 

We can also say that quantum systems are never
completely understandable with the help of a single system of classical-type 
attributes. On the other hand, in every concrete experimental context, the
subset consisting precisely of the system
attributes actualized in this context necessarily forms a classical Boolean algebra. 
The situation is somehow similar to the relation between Euclidean geometry--where the 
space is covered by a single coordinate system, and general Riemannian geometry--where 
it is only possible to cover the whole space by the atlas of local coordinate systems, each 
describing a portion of the space. 
 
Another fundamental difference between classical and
quantum mechanics is that quantum mechanics is an {\it intrinsically
stochastical} theory. In other words, probability concepts
are incorporated in the very roots of the quantum theory.
Even if we know everything about a quantum-mechanical system, 
it is still not possible to predict with certainty the outcomes of all measurements 
performed on this system. 

Accordingly, physical quantities (observables) do not possess definitive values
in quantum states. More precisely, 
if a quantum system is described by a Hilbert space $H$
then the possible states of the system are described by unit vectors 
$\psi\in H$, (and
it is assumed that unit vectors that differ up to a phase factor describe the same
physical state). Physical observables are described by selfadjoint operators $F$ acting in $H$. 
The {\it mean value} of quantum observables in states are then computed by the 
standard quantum-mechanical formula:
$$\langle F\rangle_\psi=<\psi,F\psi>.$$
 
This situation is in a sharp contrast with classical mechanics, which is 
a completely causal theory.
The states of a system are in one-to-one correspondence with points 
$\omega\in\Gamma$ of
the {\it  phase space} $\Gamma$, and physical observables are interpreted as 
the appropriate functions defined on this space. 
The values of observables $f\colon \Gamma\rightarrow \mathbb{R}$ in the states $\omega$
are simply the numbers $f(\omega)$. 
In classical mechanics, the probabilities appear as something secondary, 
whenever we do not know exactly the state of the system. A standard 
example is given by classical statistical mechanics, which studies very
complicated physical systems and so it is effectively impossible to
determine the state of the system. Accordingly in classical statistical mechanics the system is 
{\it  effectively} described by a probability distribution over the phase space
(for example--the canonical ensemble distribution).

It is natural to ask the following question---is it possible
to explain the stochasticity of quantum mechanics as a simple 
consequence of an incompleteness of quantum theory? In other words, 
as a simple consequence of the fact that quantum theory does not
include certain deeper parameters ($\Leftrightarrow$ hidden variables), which if 
included in the game would re-establish causality as in the classical 
physics.

In such a way we arive to the idea of subquantum mechanics. 

\section{Subquantum Mechanics}
By {\it  subquantum mechanics } we understand a physical theory based on the following
ideas:

\bigskip
{\bf Individual-system interpretation}

\smallskip
The theory describes {\it individual} physical systems, and goes
deeper than quantum mechanics in formulating the picture of physical reality.

\bigskip
{\bf Causality property}

\smallskip
In the framework of the theory, a concept of a 
subquantum space is defined. The elements of this space are in one-to-one correspondence
with possible subquantum states of a given physical system. It is assumed that the system
is always in some subquantum state (at the
subquantum level of description). If a subquantum state of the system is known, then the
outcomes of all quantum measurements performed on the system are completely determined. 

\bigskip
{\bf Statistical compatibility with quantum mechanics}

\smallskip
The probabilistic nature
of quantum mechanics (in other words the fact that predictions of quantum mechanics
are generally not applicable to individual physical systems, but to statictical ensembles)
is interpreted as a consequence of an incompleteness of quantum mechanics. Accordingly, 
{\it quantum states} are interpreted as certain
{\it probability measures} on the subquantum space. The probabilities of quantum
events in quantum states, coincide with the probabilities of the appropriate subquantum
counterpart-events, relative to the associated probability measures.

********\par
A theory based on the above mentioned ideas is logically possible. Classical examples
of consistent subquantum theories are Bohm-De Broglie pilot-wave theory \cite{Bo1},
Wiener-Siegel theory \cite{WS} and the theory of Bohm-Bub \cite{Bo2}. 

On the other hand, the structure of quantum mechanics implies strong restrictions
to the structure of a possible subquantum theory. These restrictions are coming from the obstacles 
contained in various so-called `no-go' theorems. Generally speaking, all 
such statements
tell us that a subquantum theory satisfying certain {\it  extra conditions } 
is not possible. Namely, as shown by von Neumann \cite{vN}, Gleason \cite{Gl} 
and Bell \cite{B2}, the concept of a subquantum state is not compatible 
with the algebra quantum observables. 
Incompatibility problems of this kind can be resolved only in the framework of 
{\it  contextual } subquantum theories \cite{S}. The common and characteristic 
property of all such theories is that the value of a given quantum observable, 
in a given subquantum state depends also on an additional entity, that is 
physically interpretable as the
corresponding {\it  measurement context. } In general, one and the same 
quantum observable realized in two different contexts will have different 
values in the same subquantum state.

The contextuality property is a necessary consistency condition. Here, we shall 
discuss quantum and subquantum mechanics within the framework of 
the {\it $C^*$-algebraic physics} (see for example \cite{BR,E,K,R}) where 
physical theories are described by the associated 
$C^*$-algebras, generated by physical observables. Physical properties are then intrinsically 
related to algebraic structure of the corresponding $C^*$-algebras of observables. In this
algebraic framework, statistical states of the system (analogs of probability measures in 
the classical context) are defined as positive normalized linear functionals 
$\rho:\Sigma\rightarrow \mathbb{C}$ on the $C^*$-algebra
$\Sigma$ of observables. For a given observable 
$F=F^*\in\Sigma$ the number $\rho(F)$ is interpreted as the {\it mean value}
of $F$ in the state $\rho$. 
A very special role is played by {\it the pure states}. The set $S(\Sigma)$ 
of all states is a convex compact (in the *-weak topology) subset of the dual space $\Sigma^*$. 
Pure states are defined as extremal elements of this convex set. 
According to Krein-Millman theorem, the set $S(\Sigma)$ is a closure of the convex linear
conbinations of pure states. Pure states are the analogs of points of the phase space in 
classical mechanics (the points viewed as maximally concentrated $\delta$-like probability 
distributions). 

The attributes of the system are in one-to-one correspondence with 
orthogonal projectors ($\Leftrightarrow$ hermitian idempotents) of the appropriate
enveloping von Neumann algebra $M(\Sigma)\supseteq\Sigma$, containing $\Sigma$ as an everywhere
dense *-subalgebra. By definition, a property $p$ holds in the state $\rho$ if and 
only if $\rho(p)=1$. The {\it negation} of the property $p$ is given by 
the projector $1-p$. A careful exposition of algebraic quantum 
mechanics, from the states-properties viewpoint, can be found in \cite{Pr}.

As far as simple quantum mechanics is concerned (systems with finitely many degrees of 
freedom, without superselection sectors) we may assume that the $C^*$-algebra 
$\Sigma$
describing the system consists of  
{\it compact operators} acting in the (infinite-dimensional separable) Hilbert state space 
$H$. The algebra $M(\Sigma)$ consists of all bounded operators in $H$.
The statistical states $\rho:\Sigma\rightarrow \mathbb{C}$
of such a system (in the sense of the above mentioned algebraic definition) are 
in a natural correspondence
with statistical operators $\hat{\rho}:H\rightarrow H$. 
The correspondence is given by the formula 
$$\rho(F)=\mathrm{Tr}(F\hat{\rho}).$$ 
Pure states then correspond to one-dimensional projectors, 
which are in a natural correspondence with unit vectors $\psi\in H$, identified 
modulo phase factors. 

\section{Contextual Extensions}

In the framework of a general approach I developed in my doctoral thesis \cite{D1}
a subquantum theory is represented by $C^*$-algebraic extensions of the form:
\begin{equation*}
0\longrightarrow \mathcal{K}\longrightarrow\widehat{\Sigma}
\xrightarrow{\mbox{$\pi$}}{} \mathrm{C}(\Omega)\longrightarrow 0
\end{equation*}
where $\widehat{\Sigma}$
is a $C^*$-algebra of `subquantum variables' and 
$\Omega$ a  compact topological
space consisting of all subquantum states of a given physical system. The elements of 
$\Omega$ are actually {\it  the characters } of $\widehat{\Sigma}$. 
By definition, characters are multiplicative
(hermitian) and non-trivial linear functionals. 
It is easy to prove that a state $\rho$ on 
a $C^*$-algebra $A$ is a character, if and only if its {\it dispersion} 
$$
\Delta_\rho(a)=\rho(a^2)-\rho(a)^2 \qquad a=a^*\in A
$$
vanishes on the elements from 
a given generating set (consisiting of hermitian elements).  

The above mentioned map $\pi$ is induced by the
evaluation on characters, in other words 
$$\pi(u)[\omega]=\omega(u).$$ 
Furthermore, the ideal 
$$\ker(\pi)=\mathcal{K}$$ 
is {\it generated} by all commutators (this is by definition {\it the commutant} 
$\mathrm{com}(\widehat{\Sigma})=\mathcal{K}$ of $\widehat{\Sigma}$).

For a given subquantum state $\omega$ and a given subquantum variable 
$u\in\widehat{\Sigma}$, the number 
$$\omega(u)=\pi(u)[\omega]$$
is interpeted as {\it  the value } of the variable $u$ in the
subquantum state $\omega$. 

The idea that the subquantum theory is deeper than quantum mechanics is incorporated
in the concept of a {\it  quantum approximation. } More precisely, this means that the 
$C^*$-algebra $\Sigma$ of quantum observables is realized as a certain approximative
image of the subquantum algebra $\widehat{\Sigma}$, with the 
help of a *-epimorphism $\phi\colon\widehat{\Sigma}\rightarrow\Sigma$
(or a more general completely positive surjection). In other words, the map is 
associating to subquantum variables their `quantum approximations'. In summary, we have
the following `double' extension diagram:
\begin{equation*}
\begin{CD}
\Bigl\{{\it Hidden Variables}\leftrightarrow\Lambda\Bigr\}\\
@VVV\\
\Bigl\{\text{\it Complementarity}\leftrightarrow\mathcal{K}\Bigr\} \longrightarrow\Bigl\{ \widehat{\Sigma}\leftrightarrow\text{\it Subq}(\Sigma) \Bigr\}
\xrightarrow{\mbox{$\pi$}}\Bigl\{ \mathrm{C}(\Omega)\leftrightarrow\text{\it Classical World}\Bigr\}\\
@VV\mbox{$\phi$}V\\
\Bigl\{\Sigma\leftrightarrow\text{\it Quantum World}\Bigr\}
\end{CD}
\end{equation*} 
where $\Lambda=\ker(\phi)$ is interpretable as the space of `truly hidden'
subquantum variables---the entities invisible at the quantum level. 

Now what about measurement contexts? The contexts are certain {\it  commutative } 
$C^*$-subalgebras of $\Sigma$. 
From the physical viewpoint, every context consists of
quantum observables that are simultaneously actualizable in a given
experimental situation (defining this context). In the conceptual framework of 
the theory, it is assumed that specifying
a quantum observable $F$ together with a measurement context $C$ defines a 
subquantum variable $u\leftrightarrow (F,C)$. 

Our subquantum theory must be statistically compatible with quantum mechanics. 
The statistical compatibility is expresed 
as the requirement that for every quantum state $\rho:\Sigma\rightarrow \mathbb{C}$
there exists a probability measure $\mu_\rho$ defined on a subquantum space $\Omega$
such that for every quantum observable $F$ and the appropriate measurement 
context $C$ we have 
$$\rho(F)= \int_{\Omega}\omega(u) d\mu_\rho(\omega) \qquad u\leftrightarrow (F,C).$$

In other words, the stochasticity of quantum states $\rho$ is interpreted as a lack of knowledge
of the elements of the subquantum space $\Omega$. 

The {\it contextuality} of the subquantum theory manifests as follows:
Let us assume that we know a quantum observable $F$, and let us assume 
that it is measured in two different contexts, corresponding to two different 
commutative $C^*$-subalgebras of $\Sigma$, say $C_1$ and $C_2$. 
From the subquantum point of view, this means that we actually 
measure two different physical quantities 
$u\leftrightarrow(F,C_1)$ and $v\leftrightarrow(F,C_2)$ 
obtained by specifying the quantum observable {\it and} the surrounding context. Generally, we
will have $u\neq v$. However a necessary consistency condition is that
$$\phi(u)=\phi(v)=F$$
because the two subquantum variables manifest as the same 
quantum observable. Moreover, 
it turns out that for every subquantum state $\omega$ there necessarily 
exist a quantum observable $F$ and contexts $C_1,C_2\ni F$ with the property: 
\begin{equation*}
\omega(u)\neq\omega(v)\qquad \begin{aligned}u&\leftrightarrow(F,C_1)\\
                        v&\leftrightarrow(F,C_2)\end{aligned}
\end{equation*}
In other words, the values of quantum observables in subquantum states
are necessarily {\it  context-dependent}. 

It is important to stress that from the subquantum viewpoint there is no contextuality at all!
The contextuality only appears if we try to understand subquantum phenomenas 
in terms of quantum observables, which are homomorphic images of the `true' 
subquantum variables.  One of the main ideas of my contextual extensions approach is to
understand contextuality as a manifestation of {\it  an inadequacy } of the
algebra $\Sigma$ of quantum observables, to describe completely physical systems. 

It is also very important to observe that probability measures $\mu_\rho$ that correspond to
quantum states $\rho$ are {\it contextually invariant} in the sense that they do not distinguish
contextual refinements of the same quantum observable. This is a trivial consequence of the
fact that there is no contextuality in quantum mechanics (and in particular quantum states 
do not know anything about contexts). In fact, it is possible to prove \cite{D1} the converse--every
contextually invariant probability distribution on $\Omega$ naturally induces a quantum state on 
$\Sigma$, assuming that $\Sigma$ is sufficiently quantum-like (for example, if $\Sigma$ is a 
von Neumann algebra without type $I_2$-factors in its central decomposition). 
This follows from a von-Neumann algebra generalization \cite{M} of the Gleason theorem \cite{Gl}. 

\section{Additional Remarks}

We saw that quantum description is understandable as an approximation of the subquantum one. 
Subquantum states of the system do not distinguish subquantum variables that coincide
modulo the elements from the commutant 
$\mathrm{com}(\widehat{\Sigma})=\mathcal{K}$ of $\widehat{\Sigma}$.
In other words, the description of a physical system in terms of subquantum space 
$\Omega$ can be understood as another aproximation of the complete description, 
obtained by ignoring all {\it  complementarity phenomenas.} 
This is because complementarity (at the subquantum level)
is based on non-commutativity of $\widehat{\Sigma}$, 
and ignoring it we arrive to a classical type theory based on 
a commutative $C^*$-algebra 
$$C(\Omega)=\widehat{\Sigma}/\mathrm{com}(\widehat{\Sigma}).$$

All phenomenas related to causality are expressible in terms of 
$\Omega$. Nevertheless, the
complete description is given by a non-commutative algebra 
$\widehat{\Sigma}$, 
not less non-commutative than the quantum one. 

One of the principal questions a coherent subquantum theory must answer is how
dynamics looks like. There exists an interesting class of subquantum theories \cite{D4} where the
subquantum space $\Omega$ is equiped with a symplectic manifold structure, such that 
the quantum evolution can be obtained from one-parametric flow generated by a smooth
function on $\Omega$, playing the role of a subquantum hamiltonian. In this sense, 
Shr\"{o}dinger equation can be viewed as a statistical version of classical Hamiltonian equations. 

The algebra $\Sigma$
of quantum observables does not admit dispersion-free states
(characters). This is a purely algebraic 
formulation of the mentioned no-go theorems, and can be viewed as a 
consequence of uncertainty relations.

It is important to mention that $C^*$-algebraic extensions considered here 
are also very interesting from the point of view of 
 non-commutative geometry. 

At first, there exist powerful techniques \cite{BDF} to apply
$C^*$-algebraic extensions (of the types similar
to our subquantum extensions)
to study topological properties of 
classical topological spaces $\Omega$. Secondly, 
noncommutative $C^*$-algebras give us 
examples of quantum spaces--the main objects of study of non-commutative 
geometry.  Let us observe that, from the non-commutative geometric viewpoint, 
subquantum states (as characters) 
play the role of {\it  points } of the quantum space associated to the algebra 
$\widehat{\Sigma}$ of subquantum variables.
 
\section{Locality \& Wholeness}
In developing a subquantum theory, it is of a crucial importance to consider the problematics
of describing the {\it composite systems}. In connection with such systems, it seems natural to
ask whether the theory possesses the following locality property:

\bigskip
{\bf Locality Property}

\smallskip
Let us consider a composite system of two distant `particles', and let
us assume that a subquantum state of the composite system is given. 
The result of an arbitrary measurement
performed on the first particle is independent of what (and if something) is 
simultaneously being
measured on the second particle---assuming there are no interactions between two 
measurements. 

\bigskip
The answer to the question of existence of a local subquantum theory is deeply connected
with the assumption about the type of probability theory applied at the subquantum level. 
Namely, if we consider a local and causal theory and assume in addition that probabilities
of all experimentally accessible events are expressible by the classical (Kolmogorovian) 
statistics, then the probabilities associated to certain joint measurements will satisfy the 
Bell inequalities \cite{B1}. On the other hand, according to quantum mechanics the same 
probablities are violating the Bell inequalities. In classical statistics, the domain of a probability 
measure is always a $\sigma$-field on the space of elementary events. 

The violation of Bell's inequalities implies that a local subquantum theory based on classical 
statistics is not possible. 

However, classical statistics is not the only way to describe the diversity of 
lack-of-knowledge situations. Moreover, from the point of view of complementarity, 
it actually seems very unnatural to use classical probability theory as the base for the 
statistics on the subquantum space. This is because:

\bigskip
\bla{i} It turns out that the family of  experimentally accessible subquantum events 
(certain subsets of subquantum space $\Omega$) is not closed under unions and 
intersections, and therefore it is not a $\sigma$-field (in contrast to classical probability theory). 

\smallskip
\bla{ii} There is no any physical justification to attach a probability to experimentally 
non-realizable events.  Accordingly, a good probability theory must operate exclusively 
with experimentally realizable events and tell us how to compute probabilities 
of such events. 

\bigskip
Starting from the above observations, it is possible to develop a {\it new probability theory} 
which includes classical probability theory as a very special case. 
And in the framework of such a generalized probability theory, it is possible to unify quantum 
mechanics with the principles of causality and locality \cite{P1,P2,G1,G2,D2,D3}. 

As pointed out in \cite{P1}, there exists a certain similarity between 
{\it probability and geometry}. Kolmogorovian probability theory corresponds to 
Euclidean geometry. However, Euclidean
geometry is not the only possible geometrical scheme. Experimental evidence, 
combined with mathematical clarity, is what 
determines suitability of geometrical systems to describe properties of 
the physical space and time. 

The same applies to probability. 

\smallskip
On the other hand, there is no experimental justification to assume locality at the 
subquantum level.  Physical reality could equally well be totally non-local.  
The idea of locality is closely related with the concepts of space and time. 
And it is not clear how the space-time looks like at the level of ultra-small distances, characterized
by the Planck lenght. Furthermore, there are various arguments to believe that the space-time
exhibits qualitatively new behaviors at the Planck scale, related to quantum fluctuations 
of geometry. Accordingly, the proper picture of the space-time should be given by non-commutative 
geometry ($\Leftrightarrow$ quantum geometry) \cite{C}. 
Quantum geometry introduces {\it a new concept of space,} by unifying classical geometry, 
non-commutative C*-algebras and basic ideas and principles of quantum physics. 

Quantum geometry deals with {\it quantum spaces}, which are very 
different from classical spaces---in particular quantum spaces may have no points at all. 
Moreover, there exist quantum spaces without any `regions'---and it seems that the 
classical concept of localizations is something very characteristic for classical spaces.

So it may happen that locality is just an illusion. At the deepest level, perhaps, everything is 
intrinsically connected with everything and the very concept of locality loses its 
standard meaning.

\end{document}